\documentclass{article_saj}
\usepackage{graphicx,saj,multicol,subeqnarray,caption}
\usepackage{natbib}
\usepackage{float}
\usepackage{comment}
\usepackage{xcolor}
\usepackage{widetext}
\usepackage{url}
\usepackage{bm}
\usepackage{tikz} 
\usepackage{pifont} 
\usepackage{amsfonts}
\usepackage{amssymb}
\usepackage{amsmath,upgreek}
\usepackage{titlesec}
\usepackage{longtable,booktabs,threeparttablex}
\titlelabel{\thetitle.\quad}
\definecolor{xlinkcolor}{cmyk}{1,0.6,0,0}
\usepackage[bookmarks=false,         
     pdfnewwindow=true,      
     colorlinks=true,   
     linkcolor=xlinkcolor,     
     citecolor=xlinkcolor,    
     filecolor=xlinkcolor,  
     urlcolor=xlinkcolor,    
final=true
]{hyperref}

\def\papertype{\ \hfill\ Original Scientific Paper}

\def\udc{52}
\def\dj{\leavevmode\setbox0=\hbox{d}\kern0pt
\rlap{\kern.215em\raise.46\ht0\hbox{-}}d}

\DeclareUnicodeCharacter{2212}{-}

\begin{document}

\parindent=.5cm
\baselineskip=3.8truemm
\columnsep=.5truecm

\newenvironment{lefteqnarray}{\arraycolsep=0pt\begin{eqnarray}}
{\end{eqnarray}\protect\aftergroup\ignorespaces}
\newenvironment{lefteqnarray*}{\arraycolsep=0pt\begin{eqnarray*}}
{\end{eqnarray*}\protect\aftergroup\ignorespaces}
\newenvironment{leftsubeqnarray}{\arraycolsep=0pt\begin{subeqnarray}}
{\end{subeqnarray}\protect\aftergroup\ignorespaces}

\markboth{\eightrm UPDATED RADIO $\Sigma − D$ RELATION AND DISTANCES TO THE SHELL-LIKE GALACTIC SNRs} 
{\eightrm Uro{\sh}evi{\cj} {\lowercase{\eightit{et al. 2025}}}}

\begin{strip}

{\ }

\vskip-1cm

\publ

\type

{\ }

\title{UPDATED RADIO $\Sigma − D$ RELATION AND DISTANCES TO THE SHELL-LIKE GALACTIC SUPERNOVA REMNANTS - IV}

\authors{D. Uro{\sh}evi{\cj}$^{1}$, B. Vukoti{\cj}$^2$, M.  An{\dj}eli{\cj}$^{1}$ and N. Mladenovi{\cj}$^{2}$}

\vskip3mm

\address{$^1$Department of Astronomy, Faculty of Mathematics,
University of Belgrade\break Studentski trg 16, 11000 Belgrade,
Serbia}

\Email{milica.andjelic@matf.bg.ac.rs, dejanu@math.rs}

\address{$^2$Astronomical Observatory, Volgina 7, 11060 Belgrade 38, Serbia}

\Email{natalija@aob.rs, bvukotic@aob.rs}

\dates{30 June, 2025}{02 September, 2025}

\summary{We present a new selected sample of $69$ Galactic supernova remnants (SNRs) for calibration of radio $\Sigma-D$ relation at $1$\,GHz. Calibrators with the most reliable distances were selected through an extensive literature search. The calibration is performed using kernel smoothing of the selected sample of calibrators in $\Sigma-D$ plane and an orthogonal offsets fitting procedure. We use the obtained calibration to derive the distances to 164 Galactic SNRs and 27 new detected SNRs/SNR candidates with none or poor distance estimates. The analysis given in this paper confirms the expected predictions from our previous papers that the kernel smoothing method is more reliable for SNR distance calibration than the orthogonal offset fitting method, {  except for the distance determinations of the very low brightness SNRs.}}


\end{strip}

\tenrm

\section{INTRODUCTION}

\indent

A relation between radio surface brightness ($\Sigma$) at frequency $\nu$ and diameter ($D$) of an expanding spherical shell like supernova remnant (SNR) was proposed by \citet{1960AZh....37..256S}:
\begin{equation}
\Sigma_\nu = A D^{\beta},    
\end{equation}
with exponent $\beta$ dependent on a spectral index $\alpha$, that characterizes  flux density distribution of radio-emission as $S_{\nu}\propto\nu^{-\alpha}$, and on change of the magnetic field strength through an SNR evolution. A parameter $A$ characterizes properties of a supernova explosion coupled with density and magnetic field strength of an SNR environment.

SNRs are strong emitters of radio synchrotron radiation. At present, there are $\approx310$ SNRs detected in the Milky Way \citep{2025JApA...46...14G}.  
The majority (92\%) of the 310 Galactic SNRs are detected at radio frequencies, while $\approx 46\%$ are detected at X-ray and $\approx 32\%$ at optical wavelengths \citep{2025JApA...46...14G}. Because of observational constraints and selection effects, there is an apparent deficit of observed number of Galactic SNRs. By using two different methods, \cite{Ranasinghe_2022} derived that there should be between $2400-5600$ SNRs in our Galaxy. While the predicted number of SNRs for the Galaxy is a few thousand, the observations/detections of new SNRs could be limited to a few hundred. With the golden era of radio astronomy at our doorstep \citep[for some of the recent SNR discoveries see][]{2025A&A...693L..15S,2024PASA...41..112F,2024MNRAS.534.2918S}, the number of radio SNRs should be significantly enlarged.

Distances to these objects can be determined using various methods, but they often require observations across the whole of the electromagnetic spectrum. When only radio flux of an SNR is available, the $\Sigma-D$ relation can be the only tool to estimate the distance to such objects. The empirical $\Sigma-D$ relation was examined in multiple works \citep{1998ApJ...504..761C,1986A&A...166..257B,2005MNRAS.360...76A,2010ApJ...719..950U,2013ApJS..204....4P,2014SerAJ.189...25P,2019SerAJ.199...23S}. $\Sigma-D$ relations are derived from the "calibrating sample" (calibrators) of shell SNRs detected at radio-wavelengths, with independently obtained distances of acceptable accuracy.

Here we present an updated sample of calibrators for the Galactic $\Sigma-D$ relation, with the most reliable independent distance estimates available in the literature (Section \ref{section_calibrators}). The method used for calibration is described in Section \ref{section_method}. Calibrations and estimated $\Sigma-D$ distances to Galactic SNRs without reliable distance estimates are presented in Section \ref{section_results} and a summary of the work is given in the last section.

\section{CALIBRATORS SELECTION AND DISTANCE UPDATES}
\label{section_calibrators}
\indent

In Table \ref{table1} we present the calibrating sample consisting of $69$ SNRs  with reliable distance estimates. For calibration of $\Sigma-D$ relation we choose SNRs whose radio-emission is non-thermal, and whose structure in radio-domain is shell-like. The starting point in our search for calibrators were catalogs by \cite{2025JApA...46...14G} and \cite{2012AdSpR..49.1313F}\footnote{http://snrcat.physics.umanitoba.ca/}, as well as the paper by \cite{Ranasinghe_2022}, in which they listed 215 SNRs (and SNR candidates) in our Galaxy, with derived distances.   {Recent studies by \cite{2023ApJS..268...61Z} and \cite{2025ApJ...988..176C} have derived or improved distances to over 150 SNRs and 170 SNR candidates.} Distances to SNRs in our calibrating sample were determined from coincidences with the observed \hbox{H\,{\sc ii}} regions and molecular clouds, or association with the red clump stars, pulsars, \hbox{H\,{\sc i}} absorption features and polarization,  H$\alpha$ line radial velocity or optical proper motion measurements. Many of distance estimates are taken from the works of \cite{Ranasinghe_2022} and \cite{Ranasinghe_2018a},   {who, for consistency, recalculated kinematic distances from literature using the values presented by \cite{2014ApJ...783..130R}. \cite{2025ApJ...988..176C} presented extinction–distance method that combines precise Gaia DR3 photometry, parallax, and stellar parameters from the SHBoost catalog \citep{2024A&A...691A..98K}  to construct extinction–distance profiles for 44 Galactic SNRs associated with  neutron star. \cite{2023ApJS..268...61Z} derived kinematic distances to SNRs that are associated with molecular clouds in the coverage of the Milky Way Imaging Scroll Painting (MWISP\footnote{http://www.radioast.nsdc.cn/mwisp.php}) project}. The list of all distance references is given below Table \ref{table1}.  

This paper is the fourth in a row of papers studying empirical $\Sigma-D$ relation by Belgrade SNR group: Paper I - \cite{2013ApJS..204....4P}; Paper II - \cite{2014SerAJ.189...25P}; Paper III - \cite{2019SerAJ.199...23S}. In Paper I the sample had 60 SNRs (55 shell-like and 5 composite) and in Paper II 65 calibrators (60 shell-like and 5 composite with a pure shell structure in radio domain). These five composite SNRs are: G11.2−0.3, G93.3+6.9 (DA 530), G189.1+3.0 (IC 443), G338.3−0.0 and G344.7−0.1. Paper III had almost twice more calibrators - 110. As  discussed in \cite{2019SerAJ.199...23S}, the list of 110 calibrators had 19 SNRs marked as objects with poor distance estimates, as well as 33 mixed-morphology SNRs. As noted in Paper III, mixed-morphology SNRs do not follow a classical SNR evolution and so can hardly be represented by the Sedov-Taylor model which is assumed in derivation of theoretical $\Sigma-D$ relation. Therefore, we did not use mixed-morphology SNRs as calibrators in this work. Additionally, the youngest Galactic SNR G1.9+0.3 which is in the flux density increasing phase of free expansion (as it was shown in \cite{2017MNRAS.468.1616P}, and in references therein) is also excluded from the set of calibrators (as it was done in Papers I, II and III).  Here, we also excluded SNRs whose distance estimates might be taken as unreliable. In the cases when independent estimates did not match within the error bars, or if the error bars were to large, we would consider that distance estimate as unreliable. In cases where multiple distances to a single SNR could be considered as reliable estimates, the average value of these distances was calculated and subsequently used to calibrate the Galactic $\Sigma-D$ relation. Distance estimates and references used for 69 calibrating SNR in this paper are presented in Table \ref{table1}. Radio fluxes and size estimates were mainly taken from \cite{2025JApA...46...14G}. For SNRs whose type presented in catalog by \cite{2025JApA...46...14G} was designated with question mark, we would visually check morphology in radio-domain, and then decide whether we would consider that object as shell (or partial-shell) or not.
For partial shell SNRs the surface brightness was calculated as measured radio flux spread over the surface as if the whole shell is observed.   {For this reason, in the list of calibrators we include only partial shells with more than 70\% shell visible in radio-image. Also, we did not use as calibrators SNRs whose geometry is highly elongated, since their surrounding ISM is probably very non-uniform, and their evolution rather complex.} At the end, instead of five composite SNRs used in Papers I, II, and III we use G11.2-0.3 and DA 530 as calibrators because of their clear separation between the shell and central field radio emission.

\onecolumn

\begin{ThreePartTable}
\begin{TableNotes}
\footnotesize
\item \textbf{References:} (1) \cite{2016ApJ...817...36S}; (2) \cite{2002AJ....124.2145V}; (3) \cite{Ranasinghe_2022}; (4) \cite{2009ApJ...694L..16H}; (5) \cite{Shan_2018}; (6) \cite{2012ApJ...753L..14H}; (7) \cite{2017RAA....17..109S}; (8) \cite{2020AJ....160..263L}; (9) \cite{Ranasinghe_2018a}; (10) \cite{Wang_2020}; (11) \cite{2021ApJS..253...17S}; (12) \cite{Ranasinghe_2018b}; (13) \cite{2018OPhyJ...4....1R}; (14) \cite{Ranasinghe_2017}; (15) \cite{2018A&A...616A..98S}; (16) \cite{2020ApJ...905...99Z}; (17) \cite{2021MNRAS.507..244F}; (18) \cite{2020ApJ...891..137Z}; (19) \cite{2012ApJ...760...25L}; (20) \cite{2022ApJ...941...17B}; (21) \cite{2002ApJ...565.1022U}; (22) \cite{2013ApJ...770..105J}; (23) \cite{2005A&A...444..871K}; (24) \cite{2007A&A...465..907T}; (25) \cite{2018MNRAS.473.1705S}; (26) \cite{2014MNRAS.441.2996A}; (27) \cite{2004ApJ...616..247Y}; (28) \cite{1993AJ....105.1060P}; (29) \cite{2010ApJ...725..894H}; (30) \cite{2006A&A...451..251L}; (31) \cite{2016ApJ...833....4Z}; (32) \cite{2013A&A...549A.107F}; (33) \cite{2016ApJ...826..108K}; (34) \cite{2007A&A...461.1013L}; (35) \cite{2015MNRAS.448.3196D}; (36) \cite{2019MNRAS.488.3129Y}; (37) \cite{2014RMxAA..50..323A}; (38) \cite{2017MNRAS.464.3029R}; (39) \cite{2015ApJ...798...82A}; (40) \cite{2019RAA....19...92S}; (41) \cite{2006MNRAS.369..416R}; (42) \cite{2004MNRAS.352.1405C}; (43) \cite{2000AJ....119..281G}; (44) \cite{1998MNRAS.296..813G}; (45) \cite{2019PASJ...71...61S}; (46) \cite{2011ApJ...742....7A}; (47) \cite{2001ApJ...551..394M}; (48) \cite{2017ApJ...851...12R}; (49) \cite{2003ApJ...585..324W}; (50) \cite{2004ApJ...604..693V}; (51) \cite{2015MNRAS.452.3470Z}; (52) \cite{1996AJ....111.1651F}; (53) \cite{2005A&A...431L...9C};  (54) \cite{2016PASJ...68S...3T}; (55) \cite{1998AJ....116.1323K}; (56) \cite{2007A&A...468..993K}; (57) \cite{2021ApJ...923...15S}; (58) \cite{2012MNRAS.421.2593T}; (59) \cite{2014ApJ...783L...2T};  (60) \cite{2007MNRAS.378.1283T}; (61) \cite{2009A&A...507..841G}; (62) \cite{2010ApJ...712..790T}; (63) \cite{2023A&A...679A.152D}; (64) \cite{2020MNRAS.493..199O}; (65) \cite{2020MNRAS.493.3947E}.
\end{TableNotes}

\begin{longtable}{@{\extracolsep{-3.0mm}} l *{10}{c} @{}}

\caption{Calibration sample for  $\Sigma-D$ relation, consisting of 69 shell SNRs, with known distances.}\\
\label{table1}\\
\toprule

No. & name & other names & $S_\mathrm{1GHz}$ & $\theta_1$ & $\theta_2$ & distance  & $D$ & $\log\Sigma$ & References\tnote{1} \\
& & & [Jy] & [arcmin] & [arcmin] & [kpc]& [pc] & [Wm$^{-2}$Hz$^{-1}$sr$^{-1}$] & \\
\midrule
\endfirsthead
\caption*{...Table 1. continued}\\
\toprule

 No. & name & other name & $S_\mathrm{1GHz}$ & $\theta_1$ & $\theta_2$ & distance  & $D$ & $\log\Sigma$ & References \\
& & & [Jy] & [arcmin] & [arcmin] & [kpc]& [pc] & [Wm$^{-2}$Hz$^{-1}$sr$^{-1}$] & \\
\endhead

\bottomrule
\endfoot

\bottomrule
\insertTableNotes  
\endlastfoot
  1 &        G4.5+6.8 &       Kepler, SN1604 &      19.0 &       3.0 &       3.0 &       5.1 &       4.5 &    -18.50 &        1 \\
  2 &        G8.7-0.1 &                  W30 &      80.0 &      45.0 &      45.0 &       3.7 &      48.4 &    -20.23 & 4, 3, 66 \\
  3 &        G9.7-0.0 &                      &       3.7 &      15.0 &      11.0 &       4.3 &      16.1 &    -20.47 & 4, 3, 66 \\
  4 &       G11.0-0.0 &                      &       1.3 &      11.0 &       9.0 &       2.7 &       7.8 &    -20.70 &    5, 66 \\
  5 & G11.2-0.3$^{*}$ &                      &      22.0 &       4.0 &       4.0 &       3.3 &       3.8 &    -18.68 &    3, 66 \\
  6 &       G12.8-0.0 &                  W33 &       0.8 &       3.0 &       4.0 &       4.3 &       4.3 &    -20.00 &    6, 66 \\
  7 &       G15.4+0.1 &                      &       5.6 &      15.0 &      14.0 &       3.0 &      12.6 &    -20.40 &       66 \\
  8 & G18.1-0.1$^{*}$ &                      &       4.6 &       8.0 &       8.0 &       6.0 &      14.0 &    -19.97 & 9, 8, 66 \\
  9 &       G18.6-0.2 &                      &       1.4 &       6.0 &       8.0 &       4.2 &       8.5 &    -20.36 &    9, 66 \\
 10 &       G18.8+0.3 &               Kes 67 &      33.0 &      17.0 &      11.0 &      13.6 &      54.1 &    -19.58 &    9, 66 \\
 11 & G21.8-0.6$^{*}$ &               Kes 69 &      65.0 &      20.0 &      20.0 &       4.7 &      27.3 &    -19.61 & 10, 9, 8, 66 \\
 12 &       G22.7-0.2 &                      &      33.0 &      26.0 &      26.0 &       4.6 &      34.8 &    -20.13 & 11, 9, 66 \\
 13 &       G23.3-0.3 &                  W41 &      70.0 &      27.0 &      27.0 &       4.8 &      37.7 &    -19.84 &    9, 66 \\
 14 &       G28.6-0.1 &                      &       5.4 &      10.8 &      10.8 &       8.9 &      28.0 &    -20.16 &   12, 66 \\
 15 &       G29.6+0.1 &                      &       0.6 &       5.0 &       5.0 &       4.7 &       6.8 &    -20.46 &   13, 66 \\
 16 &       G31.9+0.0 &                3C391 &      25.0 &       7.0 &       5.0 &       7.2 &      12.4 &    -18.97 &   14, 66 \\
 17 &       G32.4+0.1 &               Kes 32 &       0.8 &       6.0 &       6.0 &      10.7 &      18.7 &    -20.47 &    3, 66 \\
 18 &       G32.8-0.1 &               Kes 78 &      12.0 &      17.0 &      10.0 &       5.0 &      19.0 &    -19.97 &    9, 66 \\
 19 &       G33.6+0.1 &         Kes 79, HC13 &      13.5 &      10.0 &      10.0 &       4.3 &      12.5 &    -19.69 &    9, 66 \\
 20 &       G35.6-0.4 &                      &       9.0 &      10.0 &       8.0 &       3.4 &       8.8 &    -19.77 &    9, 66 \\
 21 &       G43.3-0.2 &                 W49B &      38.0 &       4.0 &       3.0 &       9.5 &       9.6 &    -18.32 &    9, 66 \\
 22 &       G54.4-0.3 &                 HC40 &      28.0 &      40.0 &      40.0 &       5.9 &      68.6 &    -20.58 &   14, 66 \\
 23 & G55.0+0.3$^{*}$ &                      &       0.6 &      20.0 &      15.0 &       4.3 &      21.7 &    -21.54 &       66 \\
 24 &       G74.0-8.5 &          Cygnus Loop &     210.0 &     200.0 &     170.0 &       0.7 &      38.9 &    -21.03 &       17 \\
 25 &       G78.2+2.1 & DR4, gamma Cygni SNR &     320.0 &      60.0 &      60.0 &       1.8 &      31.4 &    -19.87 & 5, 3, 66, 67 \\
 26 & G82.2+5.3$^{*}$ &                  W63 &     120.0 &      95.0 &      65.0 &       3.2 &      73.1 &    -20.53 & 5, 18, 66 \\
 27 & G84.2-0.8$^{*}$ &                      &      11.0 &      20.0 &      16.0 &       7.0 &      36.4 &    -20.29 &   19 , 3 \\
 28 &       G89.0+4.7 &                 HB21 &     220.0 &     120.0 &      90.0 &       1.9 &      57.4 &    -20.51 & 5, 18, 66, 67 \\
 29 & G93.7-0.2$^{*}$ &     CTB 104A, DA 551 &      65.0 &      80.0 &      80.0 &       1.9 &      44.2 &    -20.82 & 21, 10, 18 \\
 30 & G94.0+1.0$^{*}$ &              3C434.1 &      13.0 &      30.0 &      25.0 &       2.5 &      19.9 &    -20.58 & 22, 18, 3, 66 \\
 31 &      G108.2-0.6 &                      &       8.0 &      70.0 &      54.0 &       3.0 &      53.7 &    -21.50 &   24, 66 \\
 32 &      G109.1-1.0 &              CTB 109 &      20.0 &      28.0 &      28.0 &       3.1 &      25.2 &    -20.42 & 25, 66, 67 \\
 33 &      G111.7-2.1 &         Cassiopeia A &    2300.0 &       5.0 &       5.0 &       3.3 &       4.8 &    -16.86 & 26, 66, 69 \\
 34 & G114.3+0.3$^{*}$ &                      &       5.5 &      90.0 &      55.0 &       0.7 &      14.3 &    -21.78 & 27, 3, 67 \\
 35 & G116.5+1.1$^{*}$ &                      &      10.0 &      80.0 &      60.0 &       1.3 &      26.2 &    -21.50 &    27, 3 \\
 36 & G116.9+0.2$^{*}$ &                CTB 1 &       8.0 &      34.0 &      34.0 &       2.7 &      26.7 &    -20.98 & 27, 3, 66 \\
 37 & G120.1+1.4$^{*}$ &        Tycho, SN1572 &      50.0 &       8.0 &       8.0 &       3.3 &       7.7 &    -18.93 & 29, 66, 69 \\
 38 &      G127.1+0.5 &                   R5 &      12.0 &      45.0 &      45.0 &       0.6 &       7.9 &    -21.05 &   30, 66 \\
 39 &      G132.7+1.3 &                  HB3 &      45.0 &      80.0 &      60.0 &       1.9 &      38.3 &    -20.85 & 31, 66, 67 \\
 40 &      G152.4-2.1 &                      &       3.5 &     100.0 &      95.0 &       1.5 &      42.5 &    -22.26 &   32, 66 \\
 41 &      G156.2+5.7 &                      &       5.0 &     110.0 &     110.0 &       1.7 &      54.4 &    -22.21 &       33 \\
 42 &      G160.9+2.6 &                  HB9 &     110.0 &     140.0 &     120.0 &       0.8 &      30.2 &    -21.01 &   34, 67 \\
 43 &      G180.0-1.7 &                 S147 &      65.0 &     180.0 &     180.0 &       1.3 &      68.1 &    -21.52 & 35, 66, 67, 68 \\
 44 &      G190.9-2.2 &                      &       1.3 &      70.0 &      60.0 &       1.3 &      24.5 &    -22.33 &   36, 66 \\
 45 & G205.5+0.5$^{*}$ &     Monoceros Nebula &     140.0 &     220.0 &     220.0 &       1.7 &     108.8 &    -21.36 &   36, 66 \\
 46 & G206.9+2.3$^{*}$ &          PKS 0646+06 &       6.0 &      60.0 &      40.0 &       2.3 &      32.8 &    -21.42 &   37, 66 \\
 47 &      G213.0-0.6 &                      &      21.0 &     160.0 &     140.0 &       1.2 &      52.2 &    -21.85 &       36 \\
 48 & G260.4-3.4$^{*}$ &  Puppis A, MSH 08-44 &     130.0 &      60.0 &      50.0 &       1.5 &      23.9 &    -20.19 & 38, 3, 67 \\
 49 &      G290.1-0.8 &           MSH 11-61A &      42.0 &      19.0 &      14.0 &       4.8 &      22.8 &    -19.62 & 41, 3, 67 \\
 50 &      G292.2-0.5 &                      &       7.0 &      20.0 &      15.0 &       8.1 &      40.8 &    -20.45 &    42, 3 \\
 51 &      G296.8-0.3 &              1156-62 &       9.0 &      20.0 &      14.0 &       9.3 &      45.3 &    -20.32 &    44, 3 \\
 52 &      G306.3-0.9 &                      &       0.2 &       4.0 &       4.0 &      20.0 &      23.3 &    -20.82 &       45 \\
 53 &      G309.2-0.6 &                      &       7.0 &      15.0 &      12.0 &       2.8 &      10.9 &    -20.23 &       40 \\
 54 & G311.5-0.3$^{*}$ &                      &       3.7 &       4.0 &       4.0 &      12.6 &      14.7 &    -19.46 &    46, 3 \\
 55 &      G315.4-2.3 &    RCW 86, MSH 14-63 &      49.0 &      42.0 &      42.0 &       2.2 &      26.9 &    -20.38 &    40, 3 \\
 56 &      G327.4+0.4 &               Kes 27 &      25.3 &      21.0 &      21.0 &       4.3 &      26.3 &    -20.06 &    47, 3 \\
 57 &     G327.6+14.6 &               SN1006 &      19.0 &      30.0 &      30.0 &       2.0 &      17.5 &    -20.50 &   48, 49 \\
 58 &      G332.4-0.4 &              RCW 103 &      28.0 &     100.0 &     100.0 &       2.8 &      81.4 &    -21.38 &   40, 67 \\
 59 &      G332.4+0.1 &    MSH 16-51, Kes 32 &      26.0 &      15.0 &      15.0 &       7.5 &      32.7 &    -19.76 &    50, 3 \\
 60 &      G337.0-0.1 &               CTB 33 &       1.5 &       1.5 &       1.5 &      11.0 &       4.8 &    -19.00 &   52, 53 \\
 61 &      G337.2-0.7 &                      &       1.5 &       6.0 &       6.0 &       9.0 &      15.7 &    -20.20 &    54, 3 \\
 62 &      G337.8-0.1 &               Kes 41 &      15.0 &       9.0 &       6.0 &      12.3 &      26.3 &    -19.38 &    55, 3 \\
 63 &      G340.6+0.3 &                      &       5.0 &       6.0 &       6.0 &      15.0 &      26.2 &    -19.68 &       56 \\
 64 & G343.1-0.7$^{**}$ &                      &       7.8 &      27.0 &      21.0 &       4.9 &      33.9 &    -20.68 &        3 \\
 65 & G346.6-0.2$^{*}$ &                      &       8.0 &       8.0 &       8.0 &      10.4 &      24.2 &    -19.73 &    57, 3 \\
 66 &      G348.7+0.3 &              CTB 37B &      26.0 &      12.0 &      12.0 &      13.2 &      46.1 &    -19.57 &       58 \\
 67 &      G352.7-0.1 &                      &       4.0 &       8.0 &       6.0 &       7.5 &      15.1 &    -19.90 &       61 \\
 68 &      G353.6-0.7 &                      &       2.5 &      30.0 &      30.0 &       3.0 &      26.2 &    -21.38 &   62, 63 \\
 69 & G359.1-0.5$^{*}$ &                      &      14.0 &      24.0 &      24.0 &       2.7 &      18.8 &    -20.44 & 64, 65, 67 \\

 \multicolumn{10}{c}{\footnotesize {Notes: $^{*}$ SNRs with revised distances,  compared to Paper III; $^{**}$ SNRs not used as calibrators in Paper III.}}  \\

\end{longtable}
\end{ThreePartTable}

\twocolumn 
After applying these filters on SNRs listed in catalog in \cite{2025JApA...46...14G} - shell (or partial-shell) morphology   {in radio domain}; available radio-flux at 1 GHz (or estimates at other frequencies, but with known spectral index); reliable distance - our sample has 69 Galactic SNRs which were used to derive empirical $\Sigma-D$ relation.

\section{METHOD DESCRIPTION}
\label{section_method}
A two parameter straight line, fitted to the sample of calibrators is used for distance ($d$) estimates:
\begin{equation}
\log \Sigma = \log A + \beta \log D,    
\end{equation}
where $\log A$ and $\beta$ are fit parameters in the $\log \Sigma - \log D$ space. The fit is usually performed with vertical offsets, on $\log\Sigma$ axis. This fit type does not give stable results and is not consistent when using offsets on $\log D$ axis in a sense that it gives a different calibration for distance determination when compared to the case that uses $\log\Sigma$ offsets. Fits with orthogonal offsets give stable and consistent results and should be used for distance calibration \citep{2010ApJ...719..950U,2013ApJS..204....4P}.

While fitting a line through data points is a very practical way of describing the calibration, all the information contained in the calibrating sample is projected to two parameters that represent the best fit line. Given that evolution of SNRs spans over multiple orders of magnitude in surface brightness $\Sigma$ (to a lesser part also in $D$) and that SNRs expand in interstellar matter (ISM) environments of different density and consistency, the $\Sigma-D$ relation exhibits a large scatter. Some areas of $\Sigma-D$ plane may have data points that are systematically positioned away from the best fit line. Instead of a two-parameter fit line, it is more consistent to use a matrix that describes the data points density in the $\Sigma-D$ plane. This gives more consistent calibration and includes more information from the calibrating sample when compared to just using a two-parameter fit line. 

The uncertainties of orthogonal fit parameters were calculated as standard deviation from the mean values of bootstrap resamples with repetition using a random number generator \citep{SaitoMatsumoto2008}. The matrix of probability density function for data distribution in  $\Sigma-D$ plane was calculated using a Gaussian kernel. The kernel widths in both dimensions were calculated from maximum likelihood "leave one out" cross validation \citep{Duin1976}. We have updated the code from paper III \citep{2019SerAJ.199...23S} for  calculation of median value. In the previous version of the code the values for mode and median were off by a few percent on average due to the code error, which is now corrected. For each resolved value of $\log \Sigma$, from the matrix of probability density function, a median value along $\log D$ axis was used for distance calibration in order to calculate distances to objects with literature values of flux density at 1GHz ($S_\mathrm{1GHz}$) and angular diameters ($\theta_1$ and $\theta_2$). The corresponding 1GHz radio surface brightness was calculated as:

\begin{equation}
    \Sigma_\mathrm{1GHz} = \frac {S_\mathrm{1GHz}} {\Omega}= \frac {S_\mathrm{1GHz}} {\pi\theta_1\theta_2},
\end{equation}
  {where $\Omega$ is solid angle of SNR, while angular diameters $\theta_1$ and $\theta_2$ are related to distance as $\theta_1=2a\//d$, $\theta_2=2b\//d$, $a$ and $b$ being major and minor semi-axes of SNR. In this notation we calculate diameter of SNR as $D=\sqrt{\theta_1\theta_2}$}.

For presented calibrations we calculate distance fractional error for the whole calibrating sample as:
\begin{equation}
    f = \frac{100\%}{n} \times \sum_{i=1}^n\frac{\left|d_i-d_\mathrm{c}\right|}{d_i},
\end{equation}
where $d_i$ is the $i$-th SNR distance, $d_\mathrm{c}$ is the distance calculated from the calibration, while $n$ is the number of calibrators.

\section{ANALYSIS OF RESULTS: $\Sigma-D$ DISTANCES TO GALACTIC SNRs}
\label{section_results}
\indent

\begin{figure}
\centerline{\includegraphics[width=0.99\columnwidth, keepaspectratio]{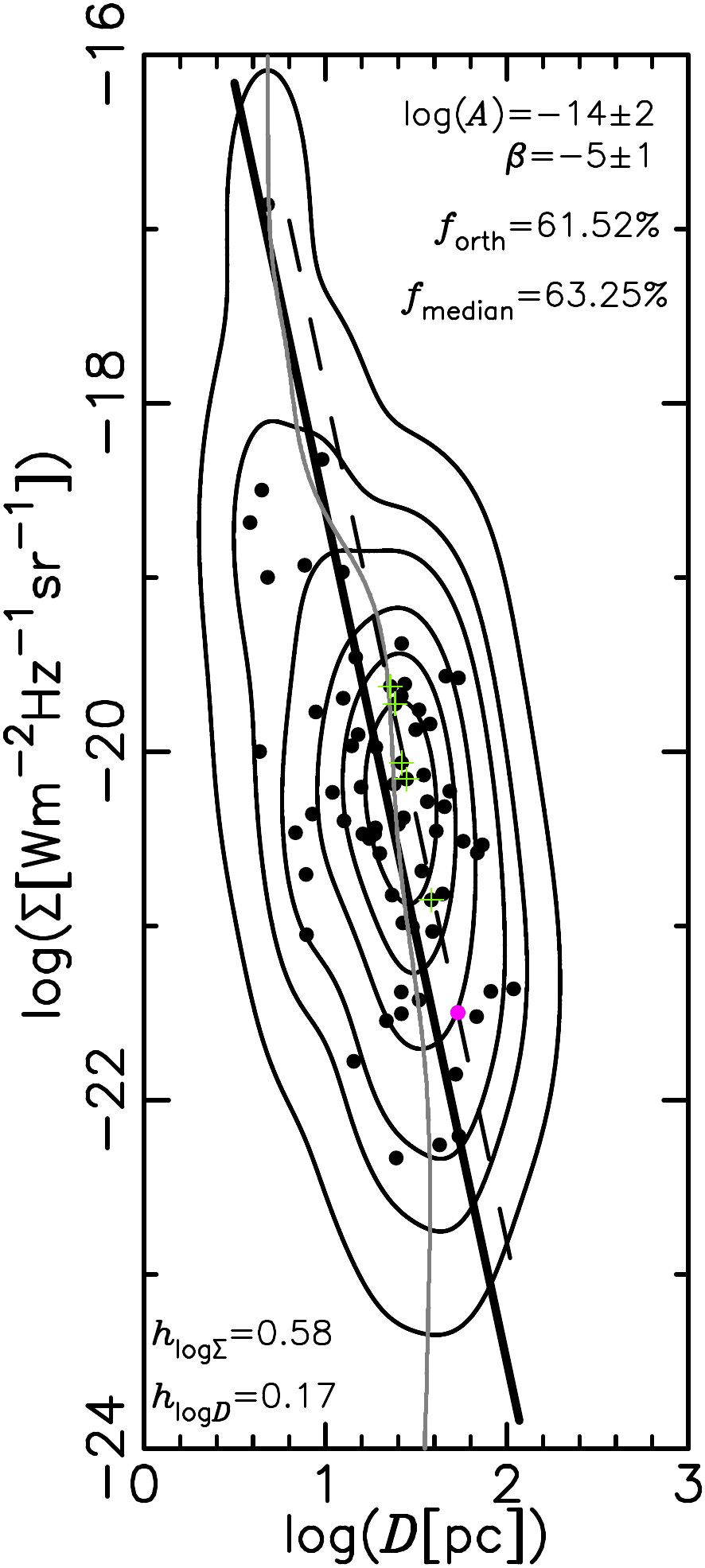}}
\caption{Calibration SNR sample consisting of $86$ calibrators with reliable distances. Black thick line presents the orthogonal offsets fit while the thin gray line designates median values of diameter distributions for the corresponding surface brightness values for the data normalized probability density matrix (contours). Contour levels are at 0.01, 0.05, 0.1, 0.2, 0.3, 0.4 and 0.5. The orthogonal fit parameters, optimal kernel widths $h$ for data probability density calculation and fractional distance errors $f$  for both, orthogonal fit and median values of diameter distributions at corresponding surface brightness values are designated on the plot. Cygnus Loop SNR is designated with a purple dot. Thin dashed line is the orthogonal fit translated along horizontal axis to the position of Cygnus Loop. Calibrators that are within $0.02$ on $\log D$ axis from the dashed line are additionally designated with green crosses. }
\label{fig1}
\end{figure}

In Fig. \ref{fig1} we present the calibrating sample and corresponding $\Sigma-D$ calibration for distance determination. The distance fractional errors are large, with similar values obtained from both methods: the orthogonal offsets fit and the kernel smoothing (calculated median value) - see Fig. \ref{fig1}. The values of the fit parameter uncertainties are indicative of the large data scatter.

Even with the most reliable distances for calibrators it is evident that $\Sigma-D$ relation has a significant intrinsic scatter which can be partly caused by distance determination uncertainties but also with the diversity of SNR environments and their parameters.

A dozen of young and bright remnants are cause of a statistically significant elongation from the rest of the sample. The majority of the sample is elliptically shaped and well grouped in $(10-100)$\,pc diameter range and $(-19, -22)$ in $\log \Sigma$ range.  This part of the sample shows a steeper $\log \Sigma-\log D$ slope, than the value of $-5$ from the orthogonal fit line. Such a steeper slope is in marginal agreement with slopes from theoretical models (which are between -6 and -4) for SNRs expanding in a uniform density ISM \citep{2018ApJ...852...84P} and is likely to be indicative of SNRs reaching the environments of the increased ISM density during their evolution \citep{2024ApJ...974..236K}. {  Moreover,
an inspection of Fig. 1 clearly shows steeper slopes in some segments of pdf line in comparison to orthogonal fit line. It can be interpreted by particular evolution and selection effects. The bump near surface brightness $10^{-19}$ W/$\mathrm {(m^2Hz\,sr)}$ probably can be explained by evolution effect of transition from the free expansion to the Sedov evolution phase. Biases in this range of the high radio surface brightnesses should not be active. On the other hand, for the obvious steeper slope of pdf line in the range of the surface brightness near and below $10^{-22}$  W/$\mathrm{(m^2Hz\,sr)}$,  the Malmquist bias should be responsible for the steepening due to missing of large diameter and low surface brightness SNRs from the sample (obvious missing of objects in the lower-right part of $\Sigma-D$ plane, right to the orthogonal fit near and below $10^{-22}$  W/$\mathrm{(m^2Hz\,sr)}$). We emphasize here that the Malmquist bias also has an influence to the steeper appearing slope of orthogonal fit line in Fig. 1. but to a lesser extent than to the pdf calibration (see also the last column of tables 2 and 3 where difference between the orthogonal fit and pdf derived distances is positive for the dimmer objects).  For this reason the orthogonal fit distances should only be used for dimmer objects with the 1 GHz brightness $\lesssim 10^{-22}$  W/$\mathrm{(m^2Hz\,sr)}$. For brighter objects the pdf method gives more realistic distances compared to distances calculated from the orthogonal fit parameters. }   

This likely contributes to a large scatter of the data in the $\Sigma-D$ plane, even when the distances are reliable and such samples are not to be described by a single straight line fit in which case the estimated distances should be less accurate than when using the data density based calibration methods.

Cygnus Loop is a shell-like SNR with the most reliable distance estimate: 725$\pm$15 pc \citep{2021MNRAS.507..244F}. Its distance is determined by using parallaxes from  {\it Gaia} mission  for stars being in front, and just behind of  the shock front. Due to this, one orthogonal fit line (thin dashed black line) in Fig. \ref{fig1} is drown through the Cygnus Loop (purple dot), as probably the best calibrator in the entire calibration sample.   {This line allows us to visually check difference between orthogonal fit through full sample, and line shifted according to one SNR with probably the best distance derived.} Additionally, calibrators that are within $0.02$ on $\log D$ axis from the dashed line are  designated with the green crosses in Fig. \ref{fig1}. These objects represent calibrators that probably share approximately the same evolutionary track with Cygnus Loop, and should be evolutionary younger "cousins" of the Cygnus Loop.   

In Table \ref{table2} we present distances to   {164} Galactic shell and partial-shell SNRs   {(catalogued in \citealt{2025JApA...46...14G})}, that do not have reliable distance estimates in literature. Distances to these SNRs were calculated using the calibrations from Fig. \ref{fig1}. We also searched literature  for recently discovered objects designated as  SNRs or SNR candidates with shell-like morphology, not present in catalog given by \cite{2025JApA...46...14G}. We found 27 these objects and they are listed in Table \ref{table3}. We give their names, surface brightness at 1GHz, angular size, spectral index and calculated distance and physical diameter. Majority of these new SNR candidates are discovered with: The H I/OH/Recombination line survey of the Milky Way -- THOR \cite{2016A&A...595A..32B}; D-configuration Very Large Array (VLA-D) continuum images of the 4–8 GHz global view on star formation (GLOSTAR) survey; LOFAR (LOw Frequency ARray)  Two-Metre Sky Survey. 

  {When using distances obtained by $\Sigma-D$ method, one should especially be careful with the interpretation in cases of partial shells or elogated SNRs, or SNRs without derived spectral index, etc. In such cases $\Sigma-D$ distance should be used as indication of probable distance.} 

\begin{table*}
\caption{Data and calculated $\Sigma-D$ distances for 164 shell SNRs that do not have their reliable distances determined by other methods. $S_\mathrm{1GHz}$ is flux density at 1 GHz, $\theta_1$ and $\theta_2$ are angular major and minor axes of SNR, $\Sigma$ is surface brightness at 1 GHz,  $d_\mathrm{orth}$ and $D_\mathrm{ort}$ are distance and physical diameter obtained from orthogonal fit calibration, while $d_\mathrm{median}$ and $D_\mathrm{median}$ are obtained  from data probability density calculation.}
\small
\parbox{\textwidth}{
\vskip.25cm
\centerline{\begin{tabular}{@{\extracolsep{-0.0mm}}ccccccccccc@{}}
\hline
\multicolumn{11}{c}{}\\
N & name & $S_\mathrm{1GHz}$ & $\theta_1$ x $\theta_2$ & $\log(\Sigma)$ & $\alpha$ & $d_\mathrm{orth}$ & $D_\mathrm{ort}$ & $d_\mathrm{median}$ & $D_\mathrm{median}$ & $\Delta d$ \\
& & [Jy] & [arcmin] & $\Sigma[\mathrm{\frac{W}{m^2\,Hz\,sr}]}$ & & [kpc] & [pc] & [kpc] & [pc] & [kpc]\\
\multicolumn{11}{c}{}\\
\hline
  1 & G0.3+0.0 & 22.00 & 15.0x8.0 & -19.56 & 0.60 & 4.9 & 16 &  6.9 & 22  & -2.0 \\
  2 & G1.0−0.1 & 15.00 & 8.0x8.0 & -19.45 & 0.60 & 6.4 & 15 &  9.2 & 21  & -2.8 \\
  3 & G1.4−0.1$^{*}$ & 2.00 & 10.0x10.0 & -20.52 & 0.55 & 8.4 & 25 &  8.7 & 25  & -0.2 \\
  4 & G3.1-0.6 & 5.00 & 52.0x28.0 & -21.29 & 0.90 & 3.2 & 35 &  2.8 & 31  & 0.4 \\
  5 & G3.7-0.2 & 2.30 & 14.0x11.0 & -20.65 & 0.65 & 7.2 & 26 &  7.2 & 26  & 0.0 \\
  6 & G3.8+0.3 & 3.00 & 18.0x18.0 & -20.86 & 0.60 & 5.5 & 29 &  5.2 & 27  & 0.3 \\
  7 & G4.2-3.5 & 3.20 & 28.0x28.0 & -21.21 & 0.60 & 4.2 & 34 &  3.7 & 30  & 0.4 \\
  8 & G4.8+6.2 & 3.00 & 18.0x18.0 & -20.86 & 0.60 & 5.5 & 29 &  5.2 & 27  & 0.3 \\
  9 & G5.2-2.6 & 2.60 & 18.0x18.0 & -20.92 & 0.60 & 5.6 & 30 &  5.3 & 28  & 0.3 \\
 10 & G5.5+0.3 & 5.50 & 15.0x12.0 & -20.34 & 0.70 & 5.8 & 22 &  6.2 & 24  & -0.5 \\
 11 & G5.9+3.1 & 3.30 & 20.0x20.0 & -20.91 & 0.40 & 5.1 & 29 &  4.7 & 28  & 0.3 \\
 12 & G6.1+0.5 & 4.50 & 18.0x12.0 & -20.50 & 0.90 & 5.7 & 24 &  5.8 & 25  & -0.2 \\
 13 & G6.4-0.1$^{*,**}$ & 48.00 & 310.0x48.0 & -21.31 & 0.55 & 1.0 & 36 &  0.9 & 31  & 0.1 \\
 14 & G6.4+4.0 & 1.30 & 31.0x31.0 & -21.69 & 0.40 & 4.7 & 43 &  3.8 & 35  & 0.9 \\
 15 & G6.5-0.4 & 27.00 & 18.0x18.0 & -19.90 & 0.60 & 3.5 & 18 &  4.4 & 23  & -0.9 \\
 16 & G7.0-0.1 & 2.50 & 15.0x15.0 & -20.78 & 0.50 & 6.3 & 28 &  6.1 & 27  & 0.2 \\
 17 & G7.2+0.2 & 2.80 & 12.0x12.0 & -20.53 & 0.60 & 7.1 & 25 &  7.2 & 25  & -0.1 \\
 18 & G7.7−3.7 & 11.00 & 22.0x22.0 & -20.47 & 0.32 & 3.7 & 24 &  3.9 & 25  & -0.1 \\
 19 & G8.3+0.0 & 1.20 & 5.0x4.0 & -20.04 & 0.65 & 15.1 & 20 &  18.1 & 23  & -3.0 \\
 20 & G8.7-5.0 & 44.00 & 26.0x26.0 & -20.01 & 0.30 & 2.5 & 19 &  3.1 & 23  & -0.5 \\
 21 & G8.9+0.4 & 9.00 & 24.0x24.0 & -20.63 & 0.60 & 3.7 & 26 &  3.7 & 26  & 0.0 \\
 22 & G9.8+0.6 & 3.90 & 12.0x12.0 & -20.39 & 0.50 & 6.6 & 23 &  7.0 & 24  & -0.4 \\
 23 & G9.9−0.8 & 6.70 & 12.0x12.0 & -20.15 & 0.40 & 5.9 & 21 &  6.8 & 24  & -0.9 \\
 24 & G11.1-0.7 & 1.00 & 11.0x7.0 & -20.71 & 0.70 & 10.5 & 27 &  10.3 & 26  & 0.2 \\
 25 & G11.1+0.1 & 2.30 & 12.0x10.0 & -20.54 & 0.40 & 7.8 & 25 &  7.9 & 25  & -0.1 \\
 26 & G11.4-0.1 & 6.00 & 8.0x8.0 & -19.85 & 0.50 & 7.7 & 18 &  9.9 & 23  & -2.2 \\
 27 & G11.8-0.2 & 0.70 & 4.0x4.0 & -20.18 & 0.30 & 18.0 & 21 &  20.5 & 24  & -2.5 \\
 28 & G12.2+0.3 & 0.80 & 6.0x5.0 & -20.40 & 0.70 & 14.5 & 23 &  15.5 & 25  & -1.0 \\
 29 & G12.7-0.0 & 0.80 & 6.0x6.0 & -20.48 & 0.80 & 13.8 & 24 &  14.3 & 25  & -0.6 \\
 30 & G13.1-0.5 & 11.00 & 38.0x28.0 & -20.81 & 0.60 & 3.0 & 28 &  2.8 & 27  & 0.1 \\
 31 & G13.5+0.2 & 3.50 & 5.0x4.0 & -19.58 & 1.00 & 12.1 & 16 &  16.9 & 22  & -4.8 \\
 32 & G14.1-0.1 & 0.50 & 6.0x5.0 & -20.60 & 0.60 & 16.0 & 25 &  16.0 & 26  & -0.0 \\
 33 & G15.1-1.6 & 5.50 & 30.0x24.0 & -20.94 & 0.00 & 3.8 & 30 &  3.6 & 28  & 0.2 \\
 34 & G15.9+0.2 & 5.00 & 7.0x5.0 & -19.67 & 0.63 & 9.5 & 16 &  13.0 & 22  & -3.5 \\
 35 & G16.0-0.5$^{**}$ & 10.00 & 2.7x15.0 & -19.43 & 0.60 & 7.9 & 15 &  11.4 & 21  & -3.5 \\
 36 & G16.2-2.7 & 2.50 & 17.0x17.0 & -20.89 & 0.40 & 5.9 & 29 &  5.5 & 27  & 0.4 \\
 37 & G17.0-0.0 & 0.50 & 5.0x5.0 & -20.52 & 0.50 & 16.9 & 25 &  17.3 & 25  & -0.4 \\
 38 & G17.4-0.1 & 0.40 & 6.0x6.0 & -20.78 & 0.70 & 15.8 & 28 &  15.2 & 27  & 0.6 \\
 39 & G17.4−2.3 & 5.00 & 24.0x24.0 & -20.88 & 0.50 & 4.2 & 29 &  3.9 & 27  & 0.2 \\
 40 & G17.8-2.6 & 5.00 & 24.0x24.0 & -20.88 & 0.50 & 4.2 & 29 &  3.9 & 27  & 0.2 \\
 41 & G17.8+16.7 & 2.70 & 51.0x45.0 & -21.75 & 0.80 & 3.1 & 44 &  2.5 & 35  & 0.6 \\
 42 & G19.1+0.2 & 10.00 & 27.0x27.0 & -20.69 & 0.50 & 3.4 & 26 &  3.3 & 26  & 0.1 \\
 43 & G21.0-0.4 & 1.10 & 9.0x7.0 & -20.58 & 0.60 & 10.9 & 25 &  11.1 & 26  & -0.1 \\
 44 & G21.6-0.8 & 1.40 & 13.0x13.0 & -20.90 & 0.50 & 7.8 & 29 &  7.3 & 28  & 0.5 \\
 45 & G21.8−3.0 & 5.00 & 60.0x60.0 & -21.68 & 0.70 & 2.4 & 42 &  2.0 & 34  & 0.5 \\
 46 & G24.7-0.6$^{**}$ & 15.00 & 8.0x30.0 & -20.03 & 0.50 & 4.3 & 19 &  5.2 & 23  & -0.9 \\
 47 & G25.1−2.3 & 8.00 & 80.0x30.0 & -21.30 & 0.50 & 2.5 & 35 &  2.2 & 31  & 0.3 \\
 48 & G27.4+0.0 & 6.00 & 4.0x4.0 & -19.25 & 0.68 & 11.6 & 13 &  16.9 & 20  & -5.3 \\
 49 & G28.3+0.2 & 1.30 & 10.0x10.0 & -20.71 & 0.70 & 9.2 & 27 &  9.0 & 26  & 0.2 \\
 50 & G28.7-0.4 & 0.90 & 9.0x9.0 & -20.78 & 0.80 & 10.6 & 28 &  10.2 & 27  & 0.4 \\
 \hline
  \multicolumn{11}{c}{Continued on next page...}\\
\end{tabular}}}
\label{table2}
\end{table*}

\begin{table*}
\parbox{\textwidth}{
\small
\vskip.25cm
\centerline{\begin{tabular}{@{\extracolsep{0.0mm}}ccccccccccc@{}} 
 \multicolumn{11}{c}{...Table 2. continued}\\
 \hline
\multicolumn{11}{c}{}\\
N & name & $S_\mathrm{1GHz}$ & $\theta_1$x$\theta_2$ & $\log(\Sigma)$ & $\alpha$ & $d_\mathrm{orth}$ & $D_\mathrm{ort}$ & $d_\mathrm{median}$ & $D_\mathrm{median}$ & $\Delta d$\\
& & [Jy] & [arcmin] & $\Sigma[\mathrm{\frac{W}{m^2\,Hz\,sr}]}$ & & [kpc] & [pc] & [kpc] & [pc] & [kpc]\\
\multicolumn{11}{c}{}\\
 \hline
 \multicolumn{11}{c}{}\\

51 & G31.5-0.6$^{*}$ & 2.00 & 18.0x18.0 & -21.03 & 0.55 & 6.0 & 31 &  5.5 & 29  & 0.5 \\
 52 & G32.0-4.9 & 22.00 & 60.0x60.0 & -21.04 & 0.50 & 1.8 & 31 &  1.6 & 29  & 0.1 \\
 53 & G34.7-0.4 & 27.00 & 240.0x35.0 & -21.32 & 0.37 & 1.3 & 36 &  1.2 & 31  & 0.2 \\
 54 & G36.6-0.7 & 1.00 & 25.0x25.0 & -21.62 & 0.70 & 5.7 & 41 &  4.7 & 34  & 1.0 \\
 55 & G36.6+2.6 & 0.70 & 17.0x13.0 & -21.32 & 0.50 & 8.3 & 36 &  7.3 & 31  & 1.0 \\
 56 & G40.5-0.5 & 11.00 & 22.0x22.0 & -20.47 & 0.40 & 3.7 & 24 &  3.9 & 25  & -0.1 \\
 57 & G41.1-0.3 & 2.50 & 25.0x4.5 & -20.48 & 0.50 & 7.8 & 24 &  8.1 & 25  & -0.3 \\
 58 & G41.5+0.4$^{*}$ & 1.00 & 10.0x10.0 & -20.82 & 0.55 & 9.7 & 28 &  9.3 & 27  & 0.4 \\
 59 & G42.0-0.1$^{*}$ & 0.50 & 8.0x8.0 & -20.93 & 0.55 & 12.8 & 30 &  11.9 & 28  & 0.9 \\
 60 & G42.8+0.6 & 3.00 & 24.0x24.0 & -21.11 & 0.50 & 4.6 & 32 &  4.2 & 29  & 0.4 \\
 61 & G43.9+1.6 & 9.00 & 60.0x60.0 & -21.42 & 0.50 & 2.1 & 38 &  1.8 & 32  & 0.3 \\
 62 & G45.7-0.4 & 4.20 & 22.0x22.0 & -20.88 & 0.40 & 4.5 & 29 &  4.3 & 27  & 0.3 \\
 63 & G46.8−0.3 & 16.00 & 15.0x15.0 & -19.97 & 0.54 & 4.3 & 19 &  5.3 & 23  & -1.0 \\
 64 & G49.2-0.7 & 160.00 & 30.0x30.0 & -19.57 & 0.30 & 1.8 & 16 &  2.5 & 22  & -0.7 \\
 65 & G51.04+0.07$^{**}$ & 3.00 & 1.8x7.5 & -19.48 & 0.55 & 14.0 & 15 &  20.1 & 21  & -6.1 \\
 66 & G53.4+0.0 & 1.50 & 10.0x10.0 & -20.65 & 0.60 & 8.9 & 26 &  8.9 & 26  & 0.0 \\
 67 & G53.6-2.2 & 8.00 & 33.0x28.0 & -20.89 & 0.50 & 3.3 & 29 &  3.1 & 27  & 0.2 \\
 68 & G55.7+3.4 & 1.00 & 23.0x23.0 & -21.55 & 0.30 & 5.9 & 40 &  5.0 & 33  & 0.9 \\
 69 & G57.2+0.8$^{**}$ & 12.00 & 1.8x12.0 & -19.08 & 0.35 & 9.2 & 12 &  12.7 & 17  & -3.5 \\
 70 & G59.5+0.1$^{*}$ & 3.00 & 15.0x15.0 & -20.70 & 0.55 & 6.1 & 27 &  6.0 & 26  & 0.1 \\
 71 & G64.5+0.9 & 0.15 & 8.0x8.0 & -21.45 & 0.50 & 16.3 & 38 &  14.0 & 33  & 2.4 \\
 72 & G65.1+0.6 & 5.50 & 90.0x50.0 & -21.74 & 0.61 & 2.2 & 43 &  1.8 & 35  & 0.4 \\
 73 & G65.3+5.7 & 42.00 & 310.0x340.0 & -22.22 & 0.60 & 0.6 & 55 &  0.4 & 37  & 0.2 \\
 74 & G67.7+1.8 & 1.00 & 15.0x12.0 & -21.08 & 0.61 & 8.2 & 32 &  7.5 & 29  & 0.7 \\
 75 & G69.7+1.0 & 2.00 & 16.0x14.0 & -20.87 & 0.70 & 6.6 & 29 &  6.3 & 27  & 0.4 \\
 76 & G73.9+0.9 & 9.00 & 27.0x27.0 & -20.73 & 0.23 & 3.4 & 27 &  3.4 & 26  & 0.1 \\
 77 & G83.0-0.3 & 1.00 & 9.0x7.0 & -20.62 & 0.40 & 11.1 & 26 &  11.1 & 26  & 0.0 \\
 78 & G93.3+6.9$^{*,**}$ & 20.00 & 9.0x27.0 & -19.91 & 0.45 & 4.0 & 18 &  5.1 & 23  & -1.1 \\
 79 & G96.0+2.0$^{*,**}$ & 26.00 & 0.3x26.0 & -18.37 & 0.60 & 10.1 & 9 &  8.9 & 8  & 1.3 \\
 80 & G114.3+0.3 & 5.50 & 90.0x55.0 & -21.78 & 0.50 & 2.2 & 44 &  1.7 & 35  & 0.4 \\
 81 & G119.5+10.2$^{**}$ & 90.00 & 36.0x90.0 & -20.38 & 0.60 & 1.4 & 23 &  1.5 & 24  & -0.1 \\
 82 & G126.2+1.6 & 6.00 & 70.0x70.0 & -21.73 & 0.50 & 2.1 & 43 &  1.7 & 35  & 0.4 \\
 83 & G166.0+4.3 & 7.00 & 55.0x35.0 & -21.26 & 0.37 & 2.7 & 35 &  2.4 & 31  & 0.3 \\
 84 & G178.2-4.2 & 2.00 & 72.0x62.0 & -22.17 & 0.50 & 2.7 & 53 &  1.9 & 37  & 0.8 \\
 85 & G179.0+2.6 & 7.00 & 70.0x70.0 & -21.67 & 0.40 & 2.1 & 42 &  1.7 & 34  & 0.4 \\
 86 & G181.1+9.5 & 0.40 & 74.0x74.0 & -22.96 & 0.40 & 3.6 & 77 &  1.7 & 37  & 1.9 \\
 87 & G182.4+4.3 & 0.50 & 50.0x50.0 & -22.52 & 0.40 & 4.3 & 63 &  2.6 & 38  & 1.7 \\
 88 & G249.5+24.5 & 27.00 & 260.0x260.0 & -22.22 & 0.70 & 0.7 & 55 &  0.5 & 37  & 0.2 \\
 89 & G261.9+5.5 & 10.00 & 40.0x30.0 & -20.90 & 0.40 & 2.9 & 29 &  2.7 & 28  & 0.2 \\
 90 & G266.2-1.2$^{**}$ & 120.00 & 50.0x120.0 & -20.52 & 0.30 & 1.1 & 25 &  1.1 & 25  & -0.0 \\
 91 & G272.3-3.2 & 0.40 & 15.0x15.0 & -21.57 & 0.60 & 9.2 & 40 &  7.7 & 34  & 1.5 \\
 92 & G279.0+1.1 & 65.00 & 30.0x95.0 & -20.46 & 0.60 & 1.5 & 24 &  1.6 & 25  & -0.1 \\
 93 & G284.3−1.8 & 11.00 & 24.0x24.0 & -20.54 & 0.30 & 3.5 & 25 &  3.6 & 25  & -0.1 \\
 94 & G288.8-6.3 & 11.00 & 108.0x96.0 & -21.80 & 0.41 & 1.5 & 45 &  1.2 & 35  & 0.3 \\
 95 & G289.7-0.3 & 6.20 & 18.0x14.0 & -20.43 & 0.20 & 5.1 & 24 &  5.3 & 25  & -0.3 \\
 96 & G294.1-0.0$^{*}$ & 2.00 & 40.0x40.0 & -21.73 & 0.55 & 3.7 & 43 &  3.0 & 35  & 0.7 \\
 97 & G296.1−0.5 & 8.00 & 37.0x25.0 & -20.89 & 0.60 & 3.3 & 29 &  3.1 & 27  & 0.2 \\
 98 & G296.5+10.0 & 65.00 & 48.0x90.0 & -20.65 & 0.50 & 1.4 & 26 &  1.4 & 26  & 0.0 \\
 99 & G296.7−0.9 & 3.00 & 15.0x8.0 & -20.42 & 0.50 & 7.4 & 23 &  7.7 & 25  & -0.4 \\
100 & G298.6-0.0 & 50.00 & 12.0x9.0 & -19.16 & 0.30 & 4.3 & 13 &  6.1 & 18  & -1.8 \\
\hline
  \multicolumn{11}{c}{Continued on next page...}\\
\end{tabular}}}

\end{table*}

\begin{table*}
\parbox{\textwidth}{
\small
\vskip.25cm
\centerline{\begin{tabular}{@{\extracolsep{0.0mm}}ccccccccccc@{}} 
 \multicolumn{11}{c}{...Table 2. continued}\\
 \hline
\multicolumn{11}{c}{}\\
N & name & $S_\mathrm{1GHz}$ & $\theta_1$x$\theta_2$ & $\log(\Sigma)$ & $\alpha$ & $d_\mathrm{orth}$ & $D_\mathrm{ort}$ & $d_\mathrm{median}$ & $D_\mathrm{median}$ & $\Delta d$ \\
& & [Jy] & [arcmin] & $\Sigma[\mathrm{\frac{W}{m^2\,Hz\,sr}]}$ & & [kpc] & [pc] & [kpc] & [pc] & [kpc] \\
\multicolumn{11}{c}{}\\
 \hline
 \multicolumn{11}{c}{}\\

101 & G299.2−2.9$^{*}$ & 0.50 & 18.0x11.0 & -21.42 & 0.55 & 9.1 & 37 &  7.9 & 32  & 1.3 \\
102 & G299.6-0.5$^{*}$ & 1.00 & 13.0x13.0 & -21.05 & 0.55 & 8.3 & 31 &  7.6 & 29  & 0.7 \\
103 & G301.4-1.0$^{*}$ & 2.10 & 37.0x23.0 & -21.43 & 0.55 & 4.4 & 38 &  3.8 & 32  & 0.6 \\
104 & G302.3+0.7 & 5.00 & 17.0x17.0 & -20.58 & 0.40 & 5.1 & 25 &  5.2 & 26  & -0.1 \\
105 & G304.6+0.1 & 14.00 & 8.0x8.0 & -19.48 & 0.50 & 6.5 & 15 &  9.2 & 21  & -2.8 \\
106 & G308.1-0.7$^{*}$ & 1.20 & 13.0x13.0 & -20.97 & 0.55 & 8.0 & 30 &  7.4 & 28  & 0.6 \\
107 & G308.4−1.4$^{*}$ & 0.50 & 12.0x6.0 & -20.98 & 0.55 & 12.3 & 30 &  11.4 & 28  & 0.9 \\
108 & G309.8+0.0 & 17.00 & 25.0x19.0 & -20.27 & 0.50 & 3.4 & 22 &  3.8 & 24  & -0.4 \\
109 & G310.6-0.3$^{*}$ & 5.00 & 8.0x8.0 & -19.93 & 0.55 & 8.0 & 19 &  10.0 & 23  & -2.0 \\
110 & G310.8-0.4$^{*}$ & 6.00 & 12.0x12.0 & -20.20 & 0.55 & 6.0 & 21 &  6.9 & 24  & -0.8 \\
111 & G312.4−0.4 & 45.00 & 38.0x38.0 & -20.33 & 0.36 & 2.0 & 22 &  2.2 & 24  & -0.2 \\
112 & G312.5-3.0$^{*}$ & 3.50 & 20.0x18.0 & -20.83 & 0.55 & 5.1 & 28 &  4.9 & 27  & 0.3 \\
113 & G315.4-0.3 & 8.00 & 24.0x13.0 & -20.41 & 0.40 & 4.5 & 23 &  4.8 & 25  & -0.3 \\
114 & G315.9-0.0$^{*}$ & 0.80 & 25.0x14.0 & -21.46 & 0.55 & 7.0 & 38 &  6.0 & 33  & 1.0 \\
115 & G316.3+0.0 & 20.00 & 29.0x14.0 & -20.13 & 0.40 & 3.5 & 20 &  4.0 & 24  & -0.6 \\
116 & G317.3-0.2$^{*}$ & 4.70 & 11.0x11.0 & -20.23 & 0.55 & 6.7 & 21 &  7.5 & 24  & -0.8 \\
117 & G318.2+0.1$^{*}$ & 3.90 & 40.0x35.0 & -21.38 & 0.55 & 3.4 & 37 &  2.9 & 32  & 0.4 \\
118 & G321.9-0.3 & 13.00 & 31.0x23.0 & -20.56 & 0.30 & 3.2 & 25 &  3.3 & 25  & -0.0 \\
119 & G321.9-1.1$^{*}$ & 3.40 & 28.0x28.0 & -21.19 & 0.55 & 4.1 & 34 &  3.7 & 30  & 0.4 \\
120 & G323.5+0.1 & 3.00 & 13.0x13.0 & -20.57 & 0.40 & 6.6 & 25 &  6.7 & 25  & -0.1 \\
121 & G327.2-0.1$^{*}$ & 0.50 & 5.0x5.0 & -20.52 & 0.55 & 16.9 & 25 &  17.3 & 25  & -0.4 \\
122 & G327.4+0.1$^{*}$ & 1.90 & 14.0x14.0 & -20.84 & 0.55 & 7.0 & 28 &  6.6 & 27  & 0.4 \\
123 & G329.7+0.4$^{*}$ & 34.00 & 40.0x33.0 & -20.41 & 0.55 & 2.2 & 23 &  2.3 & 25  & -0.1 \\
124 & G330.0+15.0 & 350.00 & 180.0x180.0 & -20.79 & 0.50 & 0.5 & 28 &  0.5 & 27  & 0.0 \\
125 & G330.2+1.0 & 5.00 & 11.0x11.0 & -20.21 & 0.30 & 6.6 & 21 &  7.5 & 24  & -0.9 \\
126 & G332.0+0.2 & 8.00 & 12.0x12.0 & -20.08 & 0.50 & 5.7 & 20 &  6.7 & 23  & -1.0 \\
127 & G332.5-5.6$^{**}$ & 35.00 & 2.0x35.0 & -19.12 & 0.70 & 5.2 & 13 &  7.4 & 18  & -2.2 \\
128 & G332.4+0.1 & 26.00 & 15.0x15.0 & -19.76 & 0.50 & 3.9 & 17 &  5.2 & 23  & -1.3 \\
129 & G335.2+0.1 & 16.00 & 21.0x21.0 & -20.26 & 0.50 & 3.6 & 22 &  4.0 & 24  & -0.4 \\
130 & G336.7+0.5 & 6.00 & 14.0x10.0 & -20.19 & 0.50 & 6.1 & 21 &  6.9 & 24  & -0.8 \\
131 & G337.3+1.0 & 15.00 & 15.0x12.0 & -19.90 & 0.55 & 4.7 & 18 &  5.9 & 23  & -1.2 \\
132 & G338.1+0.4 & 4.00 & 15.0x15.0 & -20.57 & 0.40 & 5.8 & 25 &  5.8 & 25  & -0.1 \\
133 & G340.4+0.4 & 5.00 & 10.0x7.0 & -19.97 & 0.40 & 7.8 & 19 &  9.6 & 23  & -1.8 \\
134 & G341.9-0.3 & 2.50 & 7.0x7.0 & -20.11 & 0.50 & 9.9 & 20 &  11.6 & 24  & -1.7 \\
135 & G342.0−0.2 & 3.50 & 12.0x9.0 & -20.31 & 0.40 & 7.3 & 22 &  8.0 & 24  & -0.7 \\
136 & G342.1+0.9$^{*}$ & 0.50 & 10.0x9.0 & -21.08 & 0.55 & 11.5 & 32 &  10.5 & 29  & 1.0 \\
137 & G345.1-0.2 & 1.40 & 6.0x6.0 & -20.23 & 0.70 & 12.3 & 21 &  13.7 & 24  & -1.5 \\
138 & G345.1+0.2 & 0.60 & 10.0x10.0 & -21.04 & 0.60 & 10.8 & 31 &  9.9 & 29  & 0.9 \\
139 & G345.7-0.2$^{*}$ & 0.60 & 6.0x6.0 & -20.60 & 0.55 & 14.6 & 25 &  14.6 & 26  & -0.0 \\
140 & G347.3−0.5$^{*}$ & 30.00 & 65.0x55.0 & -20.90 & 0.55 & 1.7 & 29 &  1.6 & 28  & 0.1 \\
141 & G348.5-0.0$^{**}$ & 10.00 & 10.0x10.0 & -19.82 & 0.40 & 6.1 & 18 &  7.9 & 23  & -1.8 \\
142 & G348.5+0.1 & 73.00 & 15.0x15.0 & -19.31 & 0.30 & 3.2 & 14 &  4.6 & 20  & -1.4 \\
143 & G348.8+1.1 & 0.60 & 10.0x10.0 & -21.04 & 0.70 & 10.8 & 31 &  9.9 & 29  & 0.9 \\
144 & G349.2-0.1$^{*}$ & 1.40 & 9.0x6.0 & -20.41 & 0.55 & 10.9 & 23 &  11.5 & 25  & -0.7 \\
145 & G349.7+0.2 & 2.00 & 20.0x2.5 & -20.22 & 0.50 & 10.3 & 21 &  11.7 & 24  & -1.3 \\
146 & G350.0−2.0 & 26.00 & 45.0x45.0 & -20.71 & 0.40 & 2.1 & 27 &  2.0 & 26  & 0.0 \\
147 & G350.1-0.3 & 6.00 & 4.0x4.0 & -19.25 & 0.80 & 11.6 & 13 &  16.9 & 20  & -5.3 \\
148 & G351.7+0.8 & 14.00 & 10.0x18.0 & -19.93 & 0.50 & 4.8 & 19 &  5.9 & 23  & -1.2 \\
149 & G351.9-0.9$^{*}$ & 1.80 & 12.0x9.0 & -20.60 & 0.55 & 8.4 & 25 &  8.4 & 26  & -0.0 \\
150 & G353.3-1.1 & 24.00 & 60.0x60.0 & -21.00 & 0.85 & 1.8 & 31 &  1.6 & 28  & 0.1 \\
\hline
  \multicolumn{11}{c}{Continued on next page...}\\
\end{tabular}}}

\end{table*}

\begin{table*}
\parbox{\textwidth}{
\small
\vskip.25cm
\centerline{\begin{tabular}{@{\extracolsep{0.0mm}}ccccccccccc@{}} 
 \multicolumn{11}{c}{...Table 2. continued}\\
 \hline
\multicolumn{11}{c}{}\\
N & name & $S_\mathrm{1GHz}$ & $\theta_1$x$\theta_2$ & $\log(\Sigma)$ & $\alpha$ & $d_\mathrm{orth}$ & $D_\mathrm{ort}$ & $d_\mathrm{median}$ & $D_\mathrm{median}$ & $\Delta d$ \\
& & [Jy] & [arcmin] & $\Sigma[\mathrm{\frac{W}{m^2\,Hz\,sr}]}$ & & [kpc] & [pc] & [kpc] & [pc] & [kpc] \\
\multicolumn{11}{c}{}\\
 \hline
 \multicolumn{11}{c}{}\\

151 & G353.9-2.0 & 1.00 & 13.0x13.0 & -21.05 & 0.50 & 8.3 & 31 &  7.6 & 29  & 0.7 \\
152 & G354.8−0.8$^{*}$ & 2.80 & 19.0x19.0 & -20.93 & 0.55 & 5.4 & 30 &  5.0 & 28  & 0.4 \\
153 & G355.4+0.7$^{*}$ & 5.00 & 25.0x25.0 & -20.92 & 0.55 & 4.1 & 30 &  3.8 & 28  & 0.3 \\
154 & G355.6-0.0$^{*}$ & 3.00 & 8.0x6.0 & -20.03 & 0.55 & 9.6 & 19 &  11.7 & 23  & -2.0 \\
155 & G355.9-2.5 & 8.00 & 13.0x13.0 & -20.15 & 0.50 & 5.4 & 21 &  6.3 & 24  & -0.9 \\
156 & G356.2+4.5 & 4.00 & 25.0x25.0 & -21.02 & 0.70 & 4.3 & 31 &  3.9 & 29  & 0.3 \\
157 & G356.3-0.3$^{*}$ & 3.00 & 11.0x7.0 & -20.23 & 0.55 & 8.4 & 21 &  9.4 & 24  & -1.0 \\
158 & G356.3-1.5$^{*}$ & 3.00 & 20.0x15.0 & -20.82 & 0.55 & 5.6 & 28 &  5.4 & 27  & 0.3 \\
159 & G357.7+0.3 & 10.00 & 24.0x24.0 & -20.58 & 0.40 & 3.6 & 25 &  3.7 & 26  & -0.0 \\
160 & G358.0+3.8$^{*}$ & 1.50 & 38.0x38.0 & -21.81 & 0.55 & 4.1 & 45 &  3.2 & 35  & 0.9 \\
161 & G358.1+1.0$^{*}$ & 3.00 & 20.0x20.0 & -20.95 & 0.55 & 5.2 & 30 &  4.8 & 28  & 0.4 \\
162 & G358.5-0.9$^{*}$ & 4.00 & 17.0x17.0 & -20.68 & 0.55 & 5.3 & 26 &  5.3 & 26  & 0.1 \\
163 & G359.0-0.9 & 23.00 & 23.0x23.0 & -20.18 & 0.50 & 3.1 & 21 &  3.6 & 24  & -0.4 \\
164 & G359.1+0.9$^{*}$ & 2.00 & 12.0x11.0 & -20.64 & 0.55 & 7.8 & 26 &  7.7 & 26  & 0.0 \\
\hline
 \multicolumn{11}{c}
 {\footnotesize {$^{*}$ SNRs without determined $\alpha$ and $S_\mathrm{1GHz}$. $S_\mathrm{1GHz}$ was calculated from available radio-flux and assuming $\alpha=-0.55$. \footnotesize {$^{**}$ Partial shell.}}} \\
\end{tabular}}}

\end{table*}

\begin{table*}
\caption{Data and calculated $\Sigma-D$ distances for 27 recently discovered shell SNR candidates, not cataloged in \cite{2025JApA...46...14G}, that do not have their distances determined by other methods. $d_\mathrm{orth}$ and $D_\mathrm{ort}$ are distance and physical diameter obtained from orthogonal fit calibration, while $d_\mathrm{median}$ and $D_\mathrm{median}$ are obtained  from data probability density calculation. $\Delta d$ is deference between $d_\mathrm{median}$ and $d_\mathrm{ort}$.  }
\small
\parbox{\textwidth}{
\vskip.25cm
\centerline{\begin{tabular}{@{\extracolsep{0.0mm}}cccccccccccc@{}}
\hline
\multicolumn{12}{c}{}\\
N & name & $S_\mathrm{1GHz}$ & $\theta_1$x$\theta_2$ & $\log(\Sigma)$ & $\alpha$ & $d_\mathrm{orth}$ & $D_\mathrm{ort}$ & $d_\mathrm{median}$ & $D_\mathrm{median}$& $\Delta d$ & ref. \\
& & [Jy] & [arcmin] & $\Sigma[\mathrm{\frac{W}{m^2\,Hz\,sr}]}$ & & [kpc] & [pc] & [kpc] & [pc] & [kpc] &\\
\multicolumn{12}{c}{}\\
 \hline
 \multicolumn{12}{c}{}\\
\hline
  1 & G21.8-3.0 & 5.42 & 60.0x60.0 & -21.64 & -0.72 & 2.4 & 42 &  2.0 & 34  & 0.4 &1\\
  2 & G23.11+0.18 & 5.67 & 21.7x25.0 & -20.80 & -0.63 & 4.1 & 28 &  4.0 & 27  & 0.2 &2\\
  3 & G27.06+0.04 & 1.67 & 7.5x7.5 & -20.35 & -0.53 & 10.4 & 23 &  11.1 & 24  & -0.8 &3, 4\\
  4 & G28.36+0.21 & 2.42 & 6.4x6.4 & -20.05 & -0.28 & 10.6 & 20 &  12.6 & 23  & -2.1 & 5, 6\\
  5 & G28.78−0.44 & 1.27 & 6.6x6.6 & -20.36 & -0.42 & 11.8 & 23 &  12.8 & 24  & -0.9 & 5, 6\\
  6 & G39.4−0.0$^{*}$ & 0.25 & 15.6x15.6 & -21.81 & -0.55 & 9.9 & 45 &  7.8 & 36  & 2.1 & 7\\
  7 & G39.5+0.4$^{*}$ & 0.54 & 18.0x16.8 & -21.57 & -0.55 & 7.9 & 40 &  6.7 & 34  & 1.3 &7, 4\\
  8 & G40.50+0.50 & 0.68 & 8.0x7.1 & -20.74 & -0.55 & 12.4 & 27 &  12.1 & 26  & 0.4 &8\\
  9 & G42.95−0.30 & 1.32 & 2.8x2.6 & -19.56 & -0.55 & 19.9 & 16 &  27.9 & 22  & -8.0 &8\\
 10 & G46.60+0.20 & 0.96 & 8.1x7.4 & -20.62 & -0.55 & 11.4 & 26 &  11.4 & 26  & -0.0 &8 \\
 11 & G47.78+2.02 & 0.20 & 3.7x2.8 & -20.54 & -0.55 & 26.4 & 25 &  26.9 & 25  & -0.5 &8\\
 12 & G51.26+0.11 & 14.20 & 11.3x11.3 & -19.78 & -0.40 & 5.3 & 17 &  6.9 & 23  & -1.7 &3, 4\\
 13 & G148.20+0.80 & 1.16 & 42.4x39.0 & -21.98 & -0.55 & 4.1 & 49 &  3.1 & 36  & 1.0 &8\\
 14 & G149.10+1.90 & 1.25 & 43.4x31.3 & -21.86 & -0.55 & 4.3 & 46 &  3.3 & 36  & 1.0 &8\\
 15 & G306.4+0.1$^{*}$ & 0.12 & 19.2x19.2 & -22.31 & -0.55 & 10.2 & 57 &  6.7 & 37  & 3.5 &7\\
 16 & G308.73+1.38 & 0.41 & 20.7x16.7 & -21.75 & -0.55 & 8.1 & 44 &  6.5 & 35  & 1.6 &9\\
 17 & G309.2−0.1$^{*}$ & 0.12 & 10.2x13.8 & -21.89 & -0.55 & 13.5 & 47 &  10.4 & 36  & 3.1 &7\\
 18 & G310.7-5.4$^{*}$ & 0.45 & 27.0x27.0 & -22.03 & -0.55 & 6.4 & 50 &  4.7 & 37  & 1.7 &7\\
 19 & G312.65+2.87 & 0.15 & 5.0x4.8 & -21.03 & -0.55 & 21.8 & 31 &  20.1 & 29  & 1.7 &10\\
 20 & G317.6+0.9$^{*}$ & 0.78 & 34.8x26.4 & -21.89 & -0.55 & 5.3 & 47 &  4.1 & 36  & 1.2 &6\\
 21 & G321.3-3.9 & 2.52 & 102.0x66.0 & -22.25 & -0.80 & 2.3 & 55 &  1.6 & 37  & 0.8 & 7, 11\\
 22 & G324.1+0.0 & 0.65 & 10.2x7.2 & -20.88 & -0.60 & 11.6 & 29 &  11.0 & 27  & 0.6 &7\\
 23 & G329.9−0.5 & 0.14 & 1.2x1.2 & -19.81 & -0.49 & 51.9 & 18 &  67.7 & 23  & -15.9 &12\\
 24 & G332.8−1.5$^{*}$ & 0.14 & 11.4x11.4 & -21.79 & -0.55 & 13.4 & 45 &  10.7 & 35  & 2.8 & 7\\
 25 & G333.5+0.0$^{*}$ & 1.40 & 10.8x14.4 & -20.87 & -0.55 & 8.0 & 29 &  7.5 & 27  & 0.4 &7\\
 26 & G335.7+0.9$^{*}$ & 0.12 & 12.6x12.6 & -21.94 & -0.55 & 13.1 & 48 &  9.9 & 36  & 3.2 & 7 \\
 27 & G336.8-0.6 & 0.66 & 10.8x16.8 & -21.26 & -0.60 & 8.9 & 35 &  7.9 & 31  & 1.0 &7 \\ \hline
 \hline
 \multicolumn{12}{c}{\footnotesize {Notes: $^{*}$ partial-shell SNRs; .}}  \\
\end{tabular}}}
\label{table3}
References: (1) \cite{2020MNRAS.493.2188G}; (2) \cite{2019ApJ...885..129M}; (3) \cite{2018ApJ...866...61D}; (4) \cite{2021A&A...651A..86D}; (5) \cite{2024AJ....168...42L}; (6) \cite{2023A&A...671A.145D}; (7) \cite{2025PASA...42...21M}; (8) \cite{2024A&A...690A.247T}; (9) \cite{Lazarević_2024}; (10) \cite{Unicycle}; (11) \cite{2024A&A...690A.278M}; (12) \cite{Perun}.
\end{table*}

\section{SUMMARY}
\label{section_summary}
\indent

We did a literature search for updates on reliable distances for Galactic SNRs. The calibration of $\Sigma-D$ relation was calculated using fitting of the straight line to the $\Sigma-D$ data of $69$ calibrators with reliable distance estimates,  but also with using the median values of diameter distribution for each fixed value of $\Sigma$, from the calculated probability density of the data points in the $\Sigma-D$ plane. The later method has higher fidelity in showcasing the calibrating sample, when compared to the fitting method, likely because of the discrepancy caused by the SNRs that continued their expansion into more dense regions of the ISM \citep{2024ApJ...974..236K}. This is reflected in the contour lines (Fig. \ref{fig1}) giving an impression of steeper $\Sigma-D$ slopes (on some large segments in the $\Sigma-D$ diagram) than the orthogonal fit line. These steeper slopes are in marginal agreement with the simulated $\Sigma-D$ slopes (between -4 and -6) obtained for SNRs expanding in ISM of uniform density 
\citep{2018ApJ...852...84P}. 

Due to such a slope discrepancy, methods based on data density in the $\Sigma-D$ plane should be preferred over fitting methods when calibrating the $\Sigma-D$ relation for the purpose of distance estimates, {  except for the very low surface brightness objects with 1 GHz brightness $\lesssim 10^{-22}$  W/$\mathrm{(m^2Hz\,sr)}$}. 

With the calibrations presented in Fig. \ref{fig1} and Table \ref{table1}, we have calculated distances to 164 Galactic SNRs that do not have previously determined reliable distance estimates (Table \ref{table2}). Also, in Table \ref{table3} we give distances to 27 recently discovered SNR candidates by new radio facilities, such as Australian Square Kilometer Array Pathfinder (ASKAP), Expanded Very Large Array (EVLA), The Murchison Widefield Array (MWA) and LOFAR radio interferometers.

Future research should take into account that SNRs reaching higher density ISM areas as they expand  should have different evolutionary paths in the $\Sigma-D$ plane, than the SNRs expanding in a uniform density ISM. 

\bigskip

\acknowledgements{{  We would like to thank the referee for the valuable comments and suggestions that improved the quality of this paper.}  MA and DU are supported by the Ministry of Science, Technological Development and Innovation of the Republic of Serbia, through contract no. 451-03-136/2025-03/200104. NM and BV are supported by the Ministry of Science, Technological Development and Innovation of the Republic of Serbia, through contract no. 451-03-136/2025-03/200002 made with the Astronomical Observatory (Belgrade). MA and DU are also supported through the joint project of the Serbian Academy of Sciences and Arts and Bulgarian
Academy of Sciences - "Search for optical counterparts to Galactic and extragalactic supernova remnants".}


\newcommand\eprint{in press }

\bibsep=0pt

\bibliographystyle{aa_url_saj}

\clearpage

{\ }

\newpage

\begin{strip}

{\ }

\naslov{A{\Z}URIRANA RADIO $\Sigma-D$ RELACIJA I RASTOJA{\NJ}A DO {\LJ}USKASTIH GALAKTI{\CH}KIH OSTATAKA SUPERNOVIH - {\rm IV}}

\authors{D. Uro{\sh}evi{\cj}$^{1}$, B. Vukoti{\cj}$^2$, M.  An{\dj}eli{\cj}$^{1}$ and N. Mladenovi{\cj}$^{2}$}
\vskip3mm

\address{$^1$Department of Astronomy, Faculty of Mathematics,
University of Belgrade\break Studentski trg 16, 11000 Belgrade,
Serbia}

\Email{milica.andjelic@matf.bg.ac.rs, dejanu@math.rs}

\address{$^2$Astronomical Observatory, Volgina 7, 11060 Belgrade 38, Serbia}

\Email{natalija@aob.rs, bvukotic@aob.rs}

\vskip3mm

\centerline{{\rrm UDK} \udc}

\vskip1mm

\centerline{\rit Originalni nau{\ch}ni rad}

\vskip.7cm

\baselineskip=3.8truemm

\begin{multicols}{2}

{
\rrm

U ovom radu prikazujemo  novoizdvojeni uzorak  od 69 ostataka supernovih (OSN) izabranih za kalibratore radio $\Sigma-D$ relacije na 1 {\rm GHz}. Kalibratori sa najpouzdanije odre{\dj}enim da{\lj}inama su izdvojeni posle deta{\lj}ne pretrage literature. Kalibracija je izvr{\sh}ena metodom "kernel" ravna{\nj}a izdvojenog kalibracionog uzorka u $\Sigma-D$ ravni i procedurom ortogonalnog fitova{\nj}a. Upotrebili smo dobijenu kalibraciju za odre{\dj}iva{\nj}e rastoja{\nj}a do 164 Galakti{\ch}kih OSN i do 27 novodetektovanih OSN/kandidata za OSN za koje nisu odre{\dj}ena, ili su lo{\sh}e odre{\dj}ena rastoja{\nj}a. Analiza prikazana u ovom radu potvr{\dj}uje o{\ch}ekivana predvi{\dj}a{\nj}a iz na{\sh}ih prethodnih radova da je metod "kernel" ravna{\nj}a pogodniji za odre{\dj}iva{\nj}e da{\lj}ina do OSN nego metod baziran na ortogonalnom fitova{\nj}u, osim za odre{\dj}iva{\nj}e daljina do izuzetno niskosjajnih ostataka.    

{\ }

}

\end{multicols}

\end{strip}

\end{document}